\documentclass[11pt]{article}
\usepackage{graphicx}
\usepackage{epstopdf, epsfig}
 \usepackage{amsmath}
\usepackage{natbib}
\usepackage{subcaption}

\usepackage{authblk}

\newcommand{\be}{\begin{displaymath}}
\newcommand{\bn}{\begin{equation}}
\newcommand{\bea}{\begin{eqnarray*}}
\newcommand{\eea}{\end{eqnarray*}}
\newcommand{\en}{\end{equation}}
\newcommand{\ee}{\end{displaymath}}
\newcommand{\Lpar}{L}

\newcommand{\Bm}{B_{\rm max}}

\providecommand\bnabla{\boldsymbol{\nabla}}

\title{Collisionless microinstabilities in stellarators. IV. The ion-driven trapped-electron mode}
\author[$1$]{G.G. Plunk}
\author[$2$]{J.W. Connor}
\author[$1$]{P. Helander}

\affil[$1$]{Max-Planck-Institut f\"ur Plasmaphysik, EURATOM Association, 17491 Greifswald, Germany}
\affil[$2$]{Culham Centre for Fusion Energy, Abingdon OX14 3DB, United Kingdom}

\begin{document}

\maketitle

\begin{abstract}
Optimised stellarators and other magnetic-confinement devices having the property that the average magnetic curvature is favourable for all particle orbits are called maximum-$J$ devices, and have recently been shown to be immune to trapped-particle instabilities driven by the density gradient. Gyrokinetic simulations reveal, however, that another instability can arise, which is also associated with particle trapping but causes less transport than typical trapped-electron modes. The nature of this instability is clarified here. It is shown to be similar to the ``ubiquitous mode'' in tokamaks, and is driven by ion free energy but requires trapped electrons to exist.
\end{abstract}

\section{Introduction}

Much of the transport observed in tokamaks, particularly in the plasma core, is believed to be caused by turbulence excited by ion-temperature-gradient (ITG) and trapped-electron-mode (TEM) instabilities. Less is known about transport and turbulence in stellarators, but ITG modes appear to be important, perhaps playing a role similar to that in tokamaks \citep{Watanabe-2008}. However, there are interesting and important differences between these modes in tokamaks and stellarators. For instance, the unfavourable field-line curvature driving the ``toroidal'' branch of the instability is often locally much larger in stellarators than in tokamaks  \citep{Helander-2015}, but it is very unevenly distributed over the magnetic surfaces, which has a stabilising influence \citep{Xanthopoulos-2016}. On the whole, though, it appears that the net transport caused by ITG modes can be comparable in tokamaks and stellarators. 

The situation is very different for TEMs, which can be shown to be absent in large parts of parameter space for certain types of stellarators. The collisionless TEM is driven by trapped electrons residing in regions of bad magnetic curvature, but in so-called ``maximum-$J$'' devices \citep{Rosenbluth-1968} all trapped particles experience the stabilising effect of ``good'' curvature on a time average over the orbit. For this reason, there are no collisionless, density-gradient TEMs in such configurations \citep{Proll-2012,Helander-2013}. High-beta, quasi-isodynamic stellarators \citep{Helander-2009,Nuehrenberg-2010} are, to a good approximation, maximum-$J$ devices, and gyrokinetic simulations indeed fail to find TEMs there \citep{Proll-2013}. Such modes are present in other types of stellarators, such as the Large Helical Device \citep{Nakata-2016}, but appear to be practically absent in  Wendelstein 7-X, at least in the simulations published so far. 

Instead, the simulations reveal the presence of another density-gradient-driven instability, which, in the words of \cite{Proll-2013}, ``evades standard classification''. The mode amplitude peaks in magnetic wells, as expected for TEMs, but the instability propagates in the ion diamagentic direction and draws its energy from the ions rather than the electrons. It is, however, not a traditional trapped-ion mode since the wavelength is comparable to the ion gyroradius. It is the purpose of the present paper to identify and understand the origin of this instability, which we refer to as the ion-driven trapped electron mode (ITEM). Before proceeding with the main argument for why there must be such an instability, we hasten to remark that it appears to be more benign than the conventional TEM. Nonlinear simulations so far indicate that the transport is more than an order of magnitude lower than that from TEMs in tokamaks \citep{Helander-2015}, probably because the wavelength perpendicular to the magnetic field is relatively short and the growth rate fairly small.  The latter feature, as we will show, can be traced to favourable spatial averaging of the ion drive.

This paper is the fourth part in a series on microinstabilities in stellarators (\cite{Helander-2013, Proll-2013, Plunk-2014}), and can be considered as a logical continuation of the first part.

\section{Orderings and eigenvalue problem}

We adopt two orderings that are conventional in the theory of drift-waves \citep{Helander-2013,Kadomtsev-1970}. First, in order to avoid strong Landau damping on either electrons ($e$) or ions ($i$), the phase velocity $\omega/k_\|$ along the magnetic field is taken to satisfy
	\bn v_{Ti} \ll \frac{\omega}{k_\|} \ll v_{Te}, 
	\label{drift-wave ordering}
	\en
where $v_{Ta} = (2 T_a / m_a)^{1/2}$ denotes the thermal speed of species $a$. Second, the magnetic drift frequency $\omega_{da} = {\boldsymbol k}_\perp \cdot {\boldsymbol v}_{da}$ is assumed to be much smaller than the diamagnetic frequency $\omega_{*a} = (T_a k_{\alpha}/e_a) d \ln n_a / d \psi$,
	\bn \frac{\omega_{da}}{\omega_{*a}} \ll 1, \label{small drift}
	\en
where the wave vector perpendicular to the magnetic field ${\boldsymbol B} = \nabla \psi \times \nabla \alpha$ has been written as ${\boldsymbol k}_\perp = k_\psi \nabla \psi + k_\alpha \nabla \alpha$, with $\psi$ the toroidal magnetic flux. As we shall see shortly, the effect of these two assumptions is effectively to decouple the ITG and TEM instabilities, which can otherwise seamlessly metamorphose into one another \citep{Kammerer-2008}. 

As shown by \cite{Helander-2013}, by using these assumptions it is possible to reduce the electrostatic, collisionless gyrokinetic system of equations to an eigenvalue problem involving an integral equation for the electrostatic potential $\phi(l)$ as a function of the arc length $l$ along the magnetic field,
	\bn f(\omega,l) \phi(l) = {B(l)} \int_{1/\Bm}^{1/B(l)} g(\omega,\lambda) \overline{\phi} (\lambda) 
	\frac{ d\lambda}{\sqrt{1-\lambda B}}.
	\label{integral eq2}
	\en
Here, an overbar denotes the time average over trapped-particle orbits, $\Bm$ is the maximum magnetic field strength on the flux surface in question, and we have written
	$$ f(\omega,l) = 1 + \frac{T_e}{T_i} \left[ 1 - h(\omega, l)\right], $$
	$$ g(\omega,\lambda) = \frac{1}{2} \left[1 - \frac{\omega_{*e}}{\omega} 
	+ \frac{3 \tilde \omega_{de}}{2\omega} \left(1 - \frac{(1+\eta_e) \omega_{*e}}{\omega} \right) \right], $$
	$$ h(\omega,l) = \Gamma_0(b) \left[ 1 - \frac{\omega_{*i}}{\omega} + \frac{\hat \omega_{di}}{\omega}
	- \frac{(1 + \eta_i) \omega_{*i} \hat \omega_{di}}{\omega^2} \right. $$
	$$ \left.
	+ b \left( \frac{\eta_i \omega_{*i}}{\omega} - \frac{\hat \omega_{di}}{2 \omega} 
	+ \left( 2 \eta_i + \frac{1}{2} \right) \frac{\omega_{*i} \hat \omega_{di}}{\omega^2} \right) 
	- b^2 \frac{\eta_i \omega_{*i} \hat \omega_{di}}{\omega^2} \right] $$
		$$ + b \Gamma_1(b) \left[ - \frac{\eta_i \omega_{*i}}{\omega}
	+ \frac{\hat \omega_{di}}{2 \omega} \left( 1 - \frac{\omega_{*i}}{\omega} \right) 
	- \left( \frac{3}{2} - b \right) \frac{\eta_i \omega_{*i} \hat \omega_{di}}{\omega^2} \right], $$
where $\eta_a = d \ln T_a / d \ln n$, $\Gamma_n(b) = e^{-b} I_n(b)$, $b = k_\perp^2 m_i T_i / (eB)^2$, the ion drift frequency is expressed as $\omega_{di} = \hat \omega_{di} x^2 (1 - \lambda B/2)$ and the orbit-average of the electron drift frequency as $\overline{\omega}_{de} = \tilde \omega_{de}(\lambda) x^2$, with $\lambda = v_\perp^2/(v^2B)$ and $x = v / v_{Ta}$. 

Equation (\ref{integral eq2}) admits two types of solutions, depending on whether $\phi$ vanishes at the point $l_{\rm max}$ where $B$ achieves its maximum value, $B(l_{\rm max})=\Bm$. At this point, there are no trapped particles, the right-hand side of Eq.~(\ref{integral eq2}) vanishes, and so, therefore, must either $f$ or $\phi$. In the former case, the dispersion relation is $f(\omega, l_{\rm max}) = 0$ and describes an ITG mode. In the opposite case, $\phi$ peaks somewhere in the trapped-particle region and the mode requires trapped electrons to exist. The ordering (\ref{drift-wave ordering})-(\ref{small drift}) thus allows us to discriminate between ITG modes and trapped-particle modes in an unequivocal way. 

\section{Dispersion relation}\label{disp-rel-sec}

A useful quadratic form can be obtained by multiplying Eq.~(\ref{integral eq2}) by $\phi^*/B$ and integrating along the entire field line (in ballooning space),
	\bn S[\phi,\omega] \equiv
	\int_{-\infty}^\infty f(\omega,l) |\phi|^2 \frac{dl}{B} - \int_{1/\Bm}^{1/B_{\rm min}} \sum_j \tau_j g(\omega,\lambda) | \overline{\phi}_j |^2 d\lambda = 0,
	\label{S}
	\en
where the sum is taken over all relevant magnetic wells (indexed by $j$) with magnetic field strength $B < 1/\lambda$, and 
	$$ \overline{\phi}_j(\lambda) 
	= \frac{1}{\tau_j(\lambda)} \int \frac{\phi(l) \; dl}{\sqrt{1 - \lambda B(l)}} $$
denotes the bounce average of $\phi$ over the $j$th such well, with
	$$ \tau_j(\lambda) = \int \frac{dl}{\sqrt{1 - \lambda B(l)}}. $$
Note that the integrals over $l$ are taken between bounce points defined by $\lambda B = 1$.  The form $S[\phi,\omega]$ is variational in the sense that it is stationary to first order in small variations in $\phi$ and $\omega$ satisfying the integral equation (\ref{integral eq2}) \citep{Helander-2013}. 

Equation (\ref{S}) is quadratic in $\omega$, 
	\bn
	P \omega^2 + Q \omega + R = 0,\label{dispersion eqn1}
	\en
with coefficients
	$$ P = \int_{- \infty}^\infty \left[1 + \frac{T_e}{T_i} (1-\Gamma_0) \right] \left| \phi \right|^2 \frac{dl}{B}
	- \frac{1}{2} \int_{1/\Bm}^{1/B_{\rm min}} \sum_j \tau_j \left| \overline{\phi}_j \right|^2 d\lambda, $$
	$$ Q = \frac{\omega_{\ast i} T_e}{T_i} \int_{- \infty}^\infty \left[\Gamma_0  - \eta_i b (\Gamma_0 - \Gamma_1) \right] \left| \phi \right|^2 \frac{dl}{B}
	+ \frac{\omega_{\ast e}}{2} \int_{1/\Bm}^{1/B_{\rm min}} \sum_j \tau_j \left| \overline{\phi}_j \right|^2 d\lambda, $$
	$$ R = \frac{\omega_{\ast i} T_e}{T_i} \int_{- \infty}^\infty \hat \omega_{di} \left[\Gamma_0  - \frac{b (\Gamma_0 - \Gamma_1)}{2}
	+ \eta_i  \Gamma_0 (1 - b)^2 + \eta_i b \left(\frac{3}{2} - b \right) \Gamma_1 \right] \left| \phi \right|^2 \frac{dl}{B} $$
	$$ + \frac{3 (1 + \eta_e) \omega_{\ast e}}{4} \int_{1/\Bm}^{1/B_{\rm min}} 
	\sum_j \tilde{\omega}_{dej}\tau_j \left| \overline{\phi}_j \right|^2 d\lambda, $$
where we have neglected terms that are small in $\hat \omega_{di}/\omega_{\ast i} \ll 1$. The coefficient $P$ is positive definite due to the Schwartz inequality (see \cite{Helander-2013}, where the quantity is denoted by $D[\phi]$), but $Q$ and $R$ can have either sign. Since
	\bn \omega = - \frac{1}{2 P} \left( Q \pm \sqrt{Q^2 - 4 PR} \right), 
	\label{quadratic}
	\en
it is clear that positive $R$ has a destabilising effect. We note that the first term in $R$ is proportional to the product $\omega_{\ast i} \hat \omega_{di}$, which represents the instability drive from ions due to magnetic curvature, and since the function 
	$$ \Gamma_0  - \frac{b (\Gamma_0 - \Gamma_1)}{2}
	+ \eta_i  \Gamma_0 (1 - b)^2 + \eta_i b \left(\frac{3}{2} - b \right) \Gamma_1 $$
is positive for all $b$ if $\eta_i > 0$, it follows that the ions are destabilising if $\omega_{\ast i} \hat \omega_{di} > 0$, which is the usual criterion of unfavourable magnetic curvature. The second term in $R$ involves the product $\omega_{\ast e} \tilde{\omega}_{dej}$, which represents the corresponding drive from electrons, bounce-averaged over the $j$th trapping well. In a maximum-$J$ device, this product is negative for all orbits, so that the bounce-averaged curvature is favourable for all trapped particles \citep{Rosenbluth-1968,Proll-2012,Helander-2013}. There is then no instability drive from the electrons, and conventional TEMs are absent. This was shown to be the case if $0 < \eta_e < 2/3$, independently of the orderings (\ref{drift-wave ordering})-(\ref{small drift}), by \cite{Proll-2012,Helander-2013}, and we now see it to be true for any $\eta_e > 0$ if these orderings hold, since the electron contribution to $R$ is stabilising if $\omega_{\ast e} \tilde{\omega}_{dej}$ is negative.  Furthermore, one can argue that positive $\eta_e$ is desirable in this case, since it amplifies a term that is already stabilising, making its effect even stronger.

Any instability must then be driven by the ions, but only arises if the stabilising influence of the term $Q^2$ in Eq.~(\ref{quadratic}) is small enough. According to the ordering (\ref{small drift}), $Q^2$ is larger than $PR$, but as noticed by \cite{Coppi-1977}, $Q$ must go through zero as the perpendicular wavenumber is varied, due to the behaviour of the function
 
  \bn
  F(b, \eta_i) = \Gamma_0  - \eta_i b (\Gamma_0 - \Gamma_1),\label{F-eqn}
  \en
contained in the first term. To demonstrate this, let us fix our sign conventions so that $\omega_{\ast e}$ is negative.  We observe $F$ tends to $1$ at small $b$, and the first term of $Q$ is thus positive and larger than the second term (again using the Schwartz inequality), making $Q$ positive when $b \rightarrow 0$. However, if $\eta_i > 1.64$, $F$ must become negative for $b$ greater than some value, a constant that we denote $b_c$  \citep{Helander-2013}.  Therefore, the integral quantity $Q$ can be made negative by choosing wavenumbers $k_\psi$ and $k_\alpha$ such that $b(l) > b_c$ for all $l$, and so there must be a choice of wavenumbers where $Q = 0$; note that the latter choice of wavenumbers will generally be different than the former.   Alternatively, if $\eta_i < 1.64$, we note that for $b \rightarrow \infty$,
 
	$$ \Gamma_0(b) \sim \frac{1}{\sqrt{2 \pi b}} \left( 1 + \frac{1}{8b} \right), $$
	$$ \Gamma_1(b) \sim \frac{1}{\sqrt{2 \pi b}} \left( 1 - \frac{3}{8b} \right), $$
and the first term in $Q$, although positive, becomes small. Unless $\left| \overline{\phi}_j \right| \rightarrow 0$, $Q$ will therefore again pass through zero, this time at some value of $b$ of order the inverse of the trapped-particle fraction squared, which is formally a number of order unity in our treatment. For any $\eta_i>0$ we thus expect there to be a range of wavenumbers with $b=O(1)$ in which $Q^2 < 4PR$ and instability thus prevails. In a maximum-$J$ configuration, this instability is entirely driven by ions residing in regions of bad curvature, but it requires the existence of trapped electrons. The real frequency of the mode is given by $\omega_r = -Q/(2P)$, but is limited in size by the condition on $Q^2$. The mode will thus propagate with a frequency of order $(\omega_\ast \omega_d)^{1/2}$ in either the ion or electron direction, with the direction changing sign at the value of $k_\alpha$ where $Q$ becomes zero.


\section{Iterative solution of the eigenvalue problem}\label{solns-sec}

In the previous section, we have demonstrated why the ITEM can exist even when the magnetic drift of trapped electrons is stabilising.  However, several important questions remain.  Under what conditions ({\em e.g.} in what magnetic geometries) will this mode be found?  How large will its growth rate be?  We have seen that the classical TEM is stable in special (maximum-$J$) configurations -- is it possible to devise configurations that are also immune to the ion-driven mode?

We return to the analysis of Eq.~(\ref{integral eq2}) which can be simplified by further exploiting our ordering assumptions.  We have already noted that a near cancellation of the integral quantity $Q$ is necessary for instability.  Likewise, it is necessary that the terms proportional to $\omega_{\ast}$, being formally large, must independently balance in the integral equation (\ref{integral eq2}).  Thus we can solve the integral equation iteratively.  At dominant order, we have
	\bn 
	\kappa(l) \phi(l) = {B(l)} \int_{1/\Bm}^{1/B(l)} \overline{\phi} (\lambda) \frac{ d\lambda}{\sqrt{1-\lambda B}},
	\label{integral eq2a}
	\en
where the ratio of the zeroth-order parts of $f$ and $g$ is denoted by $\kappa(l)$.  At next order we can obtain a dispersion relation for the mode frequency
	\bn
	\int_{-\infty}^{\infty} f_1(\omega, l) |\phi|^2 \frac{dl}{B} = \int_{1/\Bm}^{1/B_{\rm min}} \sum_j g_1(\omega, \lambda) \tau_j \left| \overline{\phi}_j \right|^2 d\lambda
	\label{dispersion eq2}
	\en
where we have expressed the next-order parts of $f$ and $g$ as $f_1$ and $g_1$.  These quantities include the drive terms (proportional to $\omega_{d}\omega_{\ast}$) and also contributions proportional to $\omega_{\ast}/\omega$, which are necessary for a non-zero real part of $\omega$.  The latter contributions arise mathematically by expanding $k_\alpha = k_{\alpha 0} + k_{\alpha 1}$, and $k_\psi = k_{\psi 0} + k_{\psi 1}$.  Thus, $\kappa(l)$ is equal to the quantity $2 F(b_0(l),\eta_i)$, where $F$ is defined in Eq.~(\ref{F-eqn}), and $b_0 =  |k_{\psi 0} \bnabla \psi + k_{\alpha 0} \bnabla \alpha|^2  m_i T_i / (eB)^2$.  The quantities $f_1$ and $g_1$ include those terms proportional to $k_{\alpha 1}$ or $k_{\psi 1}$ arising from the factors of $\omega_{\ast}/\omega$ contained in $f$ and $g$.  Explicit expressions for $f_1$ and $g_1$ are given in Appendix \ref{appA}.

Equation (\ref{integral eq2a}) simplifies the problem, because the eigenfunction can now be determined independently of the mode frequency, and, more importantly, we observe that it cannot depend on how the ion drift $\hat \omega_{di}$ varies with $l$.  Therefore the mode structure will not necessarily peak preferentially at locations of bad curvature (as it does {\em e.g.} for the ITG mode).  As is apparent from Eq.~(\ref{dispersion eq2}) (and also from the full expression for $R$) the ion drive contribution depends on an average of the ion magnetic drift, weighted by the mode amplitude $|\phi|^2$, and an additional factor depending on $b(l)$.  We surmise, therefore, that when there are areas of both good and bad curvature within a well (which is generally the case) there will be some cancellation due to this average, helping to stabilise the mode.  Furthermore, it is apparent that the ion drive can be eliminated if regions of good curvature outweigh the regions of bad curvature under this average.  It is therefore of primary importance to understand this equation and its solutions.

Let us discuss the general properties of Eq.~(\ref{integral eq2a}).  The integral operator on the right hand side is Hermitian.  As mentioned before, the function produced by this operator must have nodes at well endpoints where $B = \Bm$.  We also notice that the derivative of this function must be zero there too, since $B(l)$ has its maximum there.  We can make a further general observation about the symmetry of solutions to Eq.~(\ref{integral eq2a}).  Let us assume that the magnetic well is symmetric about its centre.  Then, all odd functions are in the null space of the integral operator on the right hand side, and so the right hand side must be an even function, which we may divide by $\kappa(l)$ to obtain $\phi(l)$.  We can conclude that the asymmetry of the mode in this case is induced purely by the function $\kappa(l)$, which could be controlled, to some extent, by choice of magnetic geometry.

Lastly, we note that Eq.~(\ref{integral eq2a}) suggests an interesting fundamental difference between tokamaks and stellarators.  This difference originates from the fact that ions are assumed stationary along the field line, and only the trapped part of the electron population contributes.  It is therefore only through these electrons that different points along the field line communicate, so there is no mechanism to causally connect distinct wells of height equal to the maximum $\Bm$.  In tokamaks, the well structure exactly repeats along the field line, whereas in a realistic stellarator there will be a unique value of $\Bm$ along the field line.  Thus the entire surface of the stellarator is causally connected by a finite population of trapped electrons.  Practically speaking, the numerical solution of the mode structure in a stellarator (which must be performed over a finite domain) can have a unique primary magnetic well, across which the global mode forms, coupled by trapped electrons.  However, in a tokamak, there will be a number of equivalent wells (corresponding to the number of poloidal turns spanned by the simulation domain) that are decoupled, making the notion of a single ``global mode'' somewhat artificial in this context.  This suggests that it would be sensible to limit the domain in a tokamak to a single poloidal turn for the purposes of linear TEM simulations, whereas the domain length should be longer in stellarator simulations.

\subsection{Square magnetic well}

Despite its simplicity, Eq.~(\ref{integral eq2a}) does not seem to generally admit analytical solutions.  However, there is a particularly simple limit in which a solution is readily found.  This is the case of a square magnetic well, {\it i.e.} a well of width $L$ in which the magnetic field has a constant value $B_0$, and rises abruptly to the maximum value $\Bm$ at the edges.  All trapped electrons in this well have the same bounce points (at $l = \pm L/2$), and the bounce average is just an $l$-average over the well.  We will assume $\eta_i = \eta_e = 0$ for simplicity (this has the added benefit of ensuring that $\kappa$ has no zeros).  

From Eq.~(\ref{integral eq2a}) we have
	\bn
	\kappa \phi(l) =   \frac{2 f_t}{L}\int dl' \phi(l'),\label{integral sq eq2a}
	\en
where here $\kappa = 2 \Gamma_0(b_0)$ and we introduce the trapped-particle fraction (a constant in this case),
	
	\bn
	f_t = \sqrt{1-\frac{B_0}{\Bm}}.\nonumber
	\en
The function $\phi(l)$ is obtained by simply dividing Eq.~(\ref{integral sq eq2a}) by $\kappa(l)$. We thus observe that the mode structure simply goes as $\kappa^{-1}$, and so it is purely determined by the spatial variation of $k_\perp$.  A further condition is then obtained by averaging $\phi(l)$,

	\bn
	\frac{1}{L}\int \frac{dl}{\kappa(l)} = \frac{1}{2f_t}.\label{sq-eigenvalue-eqn}
	\en
This condition determines allowable values of $k_{\alpha 0}$ and $k_{\psi 0}$.  For simplicity, we take $k_{\alpha1} = 0$, and from Eq.~(\ref{dispersion eq2}) we obtain a dispersion relation for a purely growing (or decaying) mode

	\bn
	\frac{4 f_t^2}{L}\int \frac{dl}{\kappa^2} \left[1 + \frac{T_e}{T_i}(1- h_1) \right]   =  B_0\int_{1/\Bm}^{1/B_0} \frac{g_1 d\lambda}{\sqrt{1-\lambda B_0}}
	\label{dispersion sq eq2}
	\en
where

	\bn
	g_1(\omega, \lambda) = \frac{1}{2}\left[ 1 - \frac{3\omega_{*e} \tilde \omega_{de}}{2\omega^2} \right],\nonumber
	\en
and

	$$h_1(\omega, l) = \Gamma_0(b) + \left(\frac{\omega_{*i} \hat \omega_{di}}{\omega^2}\right) \left[\Gamma_0(b) \left(\frac{b}{2}  - 1 \right) - \frac{b}{2} \Gamma_1(b)\right].\nonumber$$
Note that Eq.~(\ref{dispersion sq eq2}) is quadratic in $\omega$, with no linear term, and thus it follows that, assuming electrons are stabilising ($\omega_{*e} \tilde \omega_{de} < 0$), instability requires that the condition

	$$\frac{1}{L}\int  \frac{dl\; \omega_{*i} \hat \omega_{di}}{\kappa^2}\left[\Gamma_0(b) \left(1 - \frac{b}{2} \right) + \frac{b}{2} \Gamma_1(b)\right] > 0,$$
be satisfied.  Note that the quantity within the brackets is positive definite, and so instability requires a kind of ``average-bad'' curvature, where the weight of the average depends on $l$ only via $b(l)$.  As a further simplification, we can consider a system with linear magnetic shear, such that $\bnabla \alpha = (\hat{\boldsymbol y} +  \hat{\boldsymbol x} s l) |\bnabla \alpha|_{l = 0}$, $\bnabla \psi = \hat{\boldsymbol x} |\bnabla \psi|_{l = 0}$, where we have introduced $s$, the inverse magnetic shear length.  Defining $k_x = k_\psi |\bnabla \psi|_{l = 0} $, $k_y = k_\alpha |\bnabla \psi|_{l = 0}$, and $l_0 = -k_x/(k_y s)$, we obtain $k_\perp^2 = k_y^2(s^2(l - l_0)^2 + 1)$.  Assuming that the location $l_0$ is outside the magnetic well, at a sufficient distance, and that $s$ and $k_y$ are positive, we obtain $k_\perp \approx k_y | l - l_0| s$.  Then, taking $b \gg 1$, we can use the large-argument expansions of $\Gamma_0$ and $\Gamma_0$ to obtain

\bn
\frac{1}{\kappa^2}\left[\Gamma_0(b) \left(1 - \frac{b}{2} \right) + \frac{b}{2} \Gamma_1(b)\right] \approx \frac{3}{8}\sqrt{\frac{\pi}{2}} k_{y0} \rho |l - l_0| s,
\en
Expanding Eq.~(\ref{sq-eigenvalue-eqn}) for $b \gg 1$ yields the solution for $k_{y0}$,
\bn
k_{y0} \rho = \frac{1}{\sqrt{2 \pi} |l_0| s f_t}.
\en

As a second simplification, we can take $\tilde \omega_{de}$ to be independent of pitch angle $\lambda$, which is a good approximation in some geometries \citep{kesner-hastie}.  Then the integral on the right hand side of Eq.~(\ref{dispersion sq eq2}) simply yields another factor of $2 f_t$.  Assuming now that electrons are stabilising, $\omega_{*e} \tilde \omega_{de} < 0$, it is easy to devise a function $\omega_{di}(l)$ (involving areas of both good and bad curvature) that yields a stable mode.  The strategy, which is simply to choose the values of good curvature to reside in regions that are more strongly weighted by the function $|l-l_0|$, may not work for all values of $l_0$, so it still seems challenging to obtain absolute stability of the ITEM mode, given that some areas of bad curvature must be present for a magnetic field line that traces out a topological torus.  Still, the possibility cannot be ruled out entirely.

\subsection{Numerical solutions}

Let us next consider numerical solutions of Eq.~(\ref{integral eq2a}).  A uniform grid is used in $l$-space, and the integrals in $l$ and $\lambda$ are evaluated numerically.  We can approach Eq.~(\ref{integral eq2a}) as an eigenvalue equation, considering $k_\psi$ as a fixed parameter.  We take the locations of maximum magnetic field to be $\pm \Lpar/2$, normalize $\hat{l} = l/\Lpar$, and decompose $\kappa(l) = \nu_i \hat{\kappa}(\hat{l}) $ where we choose $\hat{\kappa}(0) = 1$, and $\nu_i$ denotes the eigenvalue, with index $i$.  Two choices, $\hat{\kappa} = 1/|\hat{l}-1|$, and $\kappa = 1$ are considered, corresponding, respectively, to linear magnetic shear with $b \gg 1$, and zero magnetic shear.  Thus, the allowable values of $k_{\alpha 0}$ are determined by the eigenvalues $\nu_i$.  The mode frequency $\omega$ can then immediately be determined by Eq.~(\ref{dispersion eq2}) (which is merely a quadratic dispersion relation) where $k_{\alpha 1}$ is a free parameter whose sign will determine the direction of mode propagation.  Note that the choice $k_{\alpha 1} = 0$ causes the quadratic form $Q$ to be precisely zero, and the corresponding mode must then have zero real frequency.  

\begin{figure*}
    \centering
    \begin{subfigure}[b]{0.5 \textwidth}
        \centering
	\includegraphics[width=\textwidth]{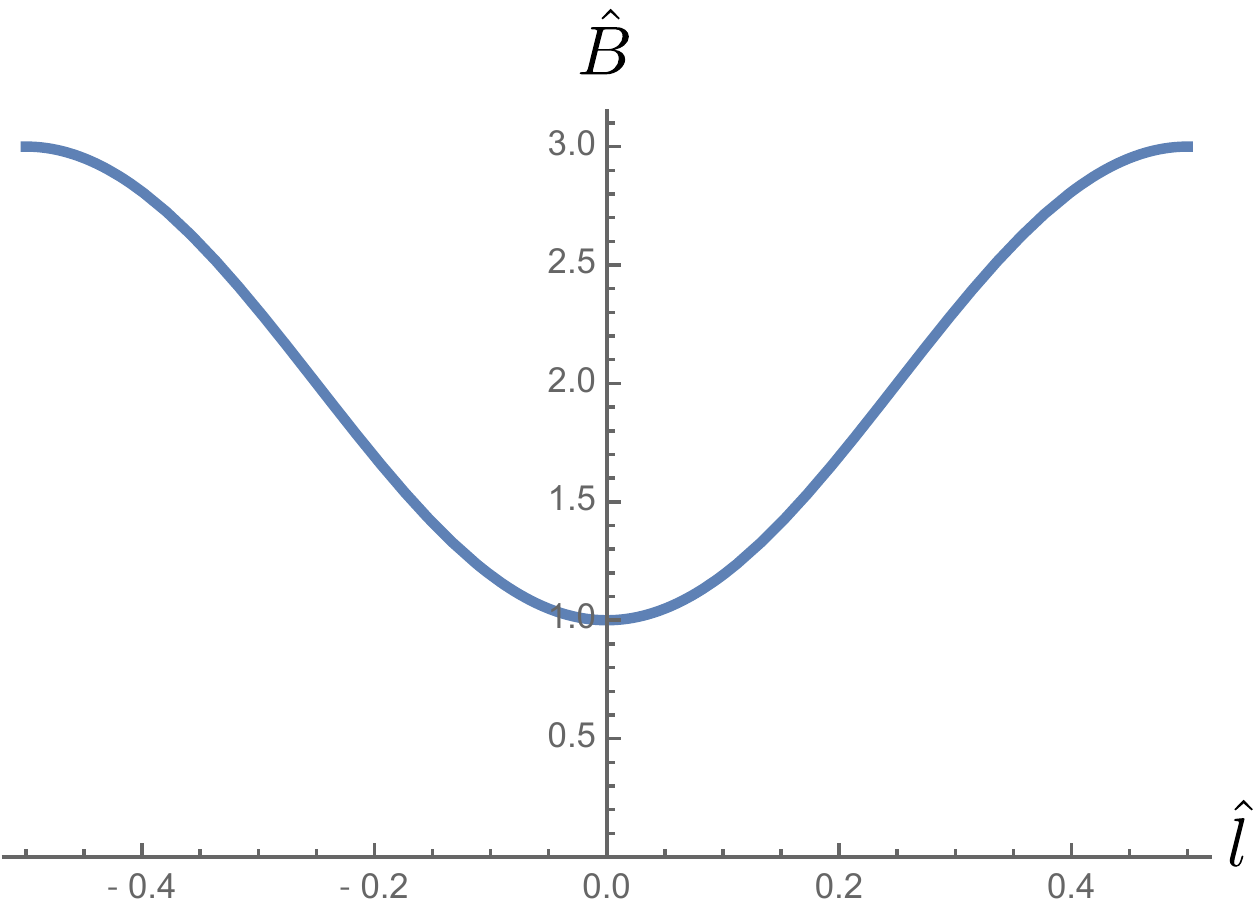}
        \caption{$\hat{B} = 2 - \cos(2\pi \hat{l})$.}
        \label{well-sin-fig}
    \end{subfigure}%
    ~ 
    \begin{subfigure}[b]{0.5 \textwidth}
        \centering
	\includegraphics[width=\textwidth]{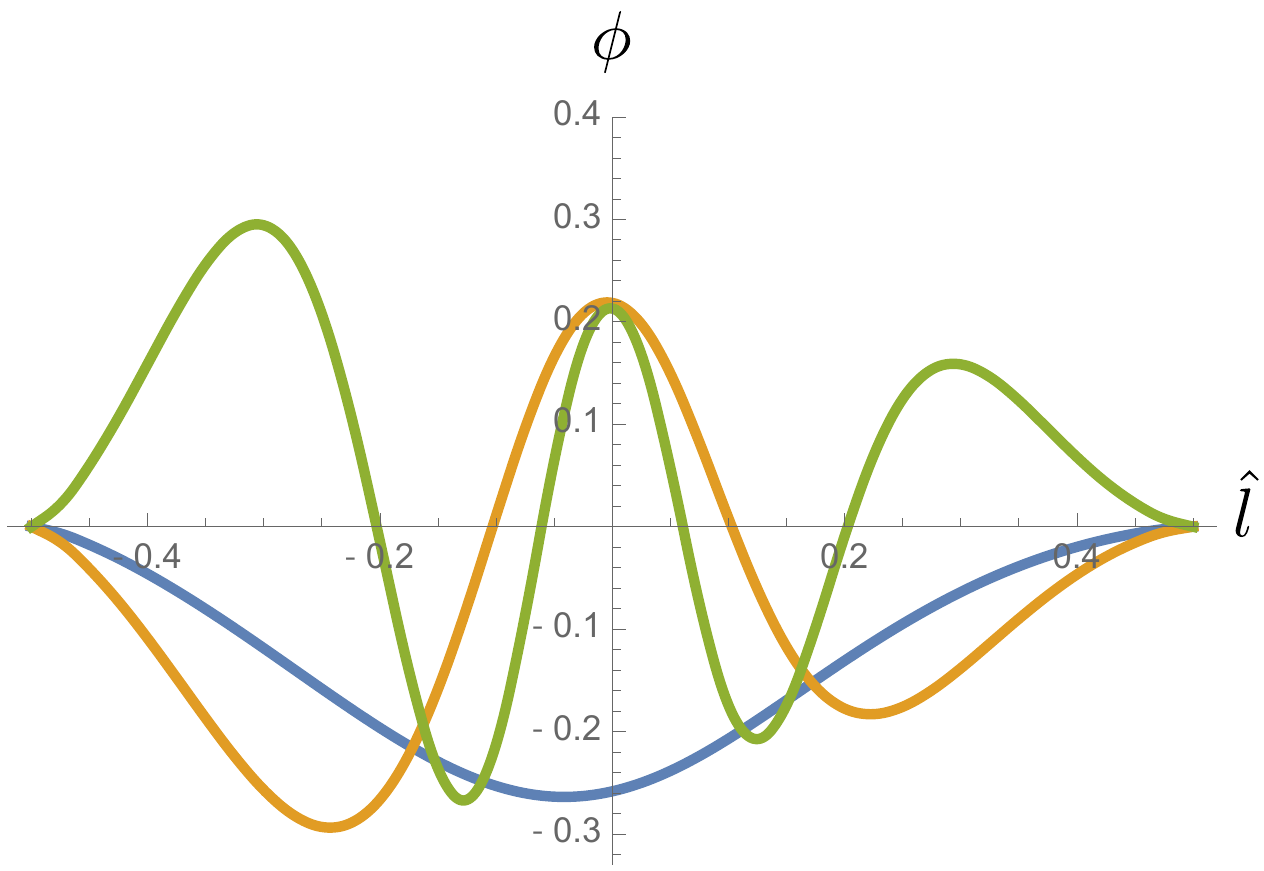}
        \caption{Eigenvalues: $1.47, 0.64, 0.42$.}
	\label{eigenmode-fig}
    \end{subfigure}
    \newline
    \begin{subfigure}[b]{0.5 \textwidth}
        \centering
	\includegraphics[width=\textwidth]{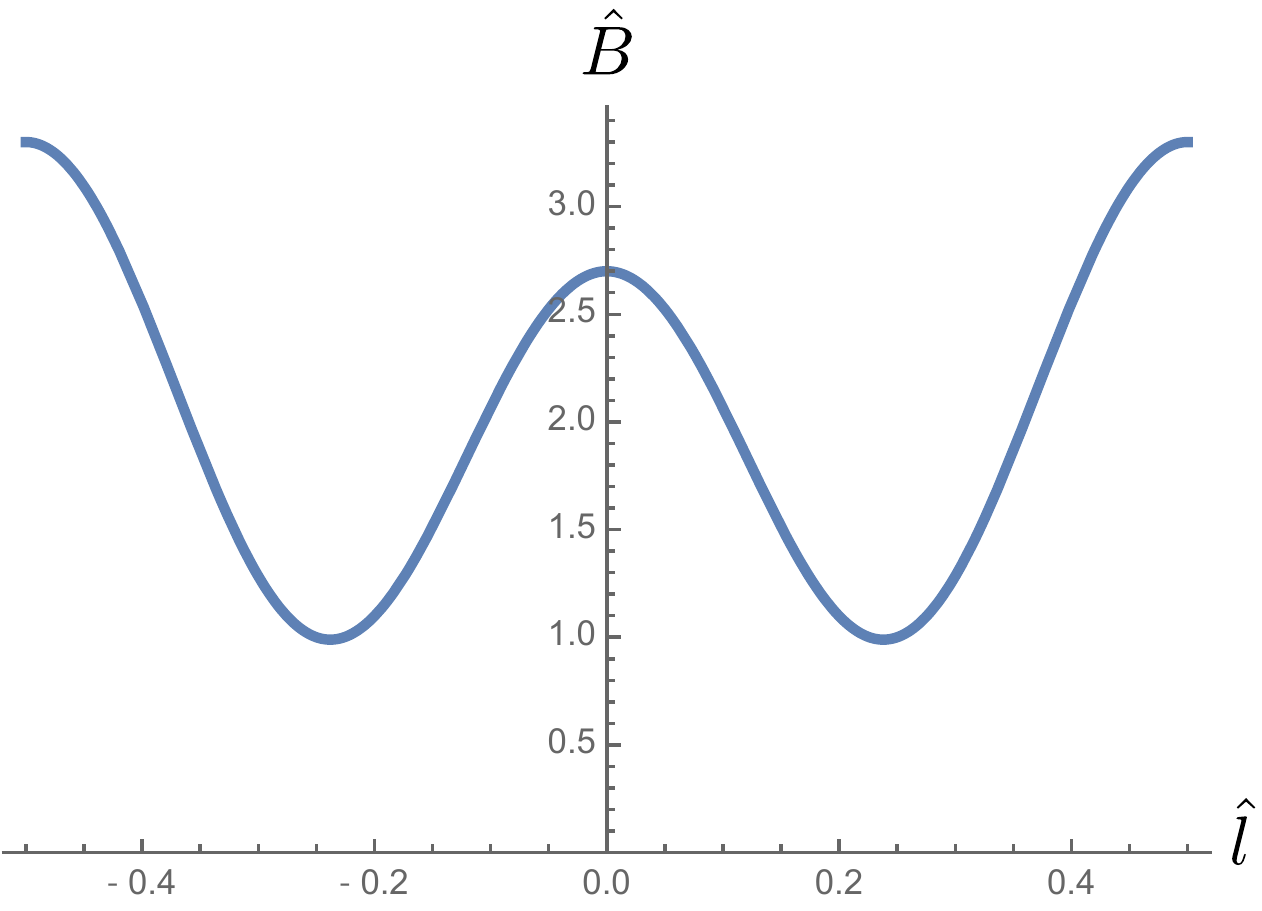}
        \caption{$\hat{B} = 2 - 0.3 \cos(2\pi \hat{l}) + \cos(4 \pi \hat{l})$.}
        \label{well-db-fig}
    \end{subfigure}%
    ~ 
    \begin{subfigure}[b]{0.5 \textwidth}
        \centering
	\includegraphics[width=\textwidth]{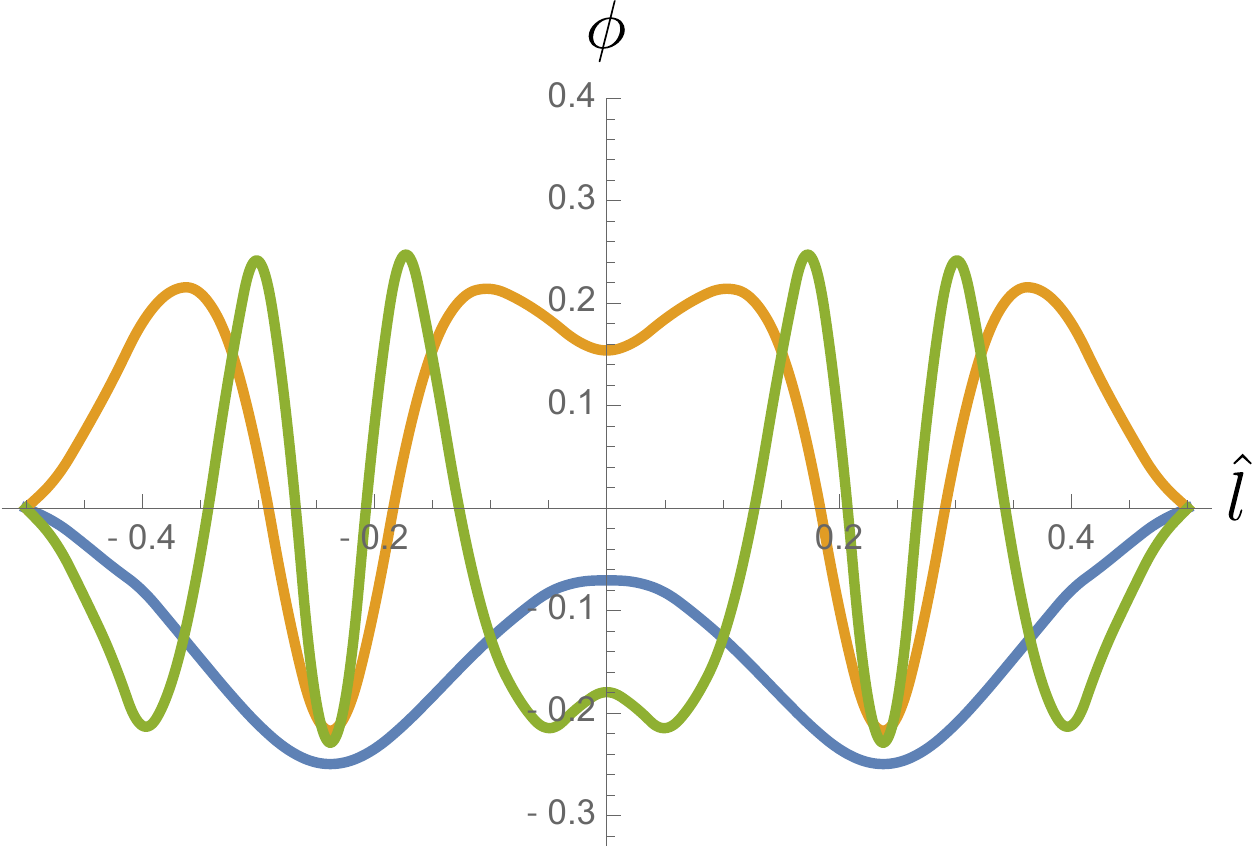}
        \caption{Eigenvalues: $1.49, 0.67, 0.46$.}
	\label{eigenmode-db-fig}
    \end{subfigure}
    \newline
    \begin{subfigure}[b]{0.5 \textwidth}
        \centering
	\includegraphics[width=1.0\textwidth]{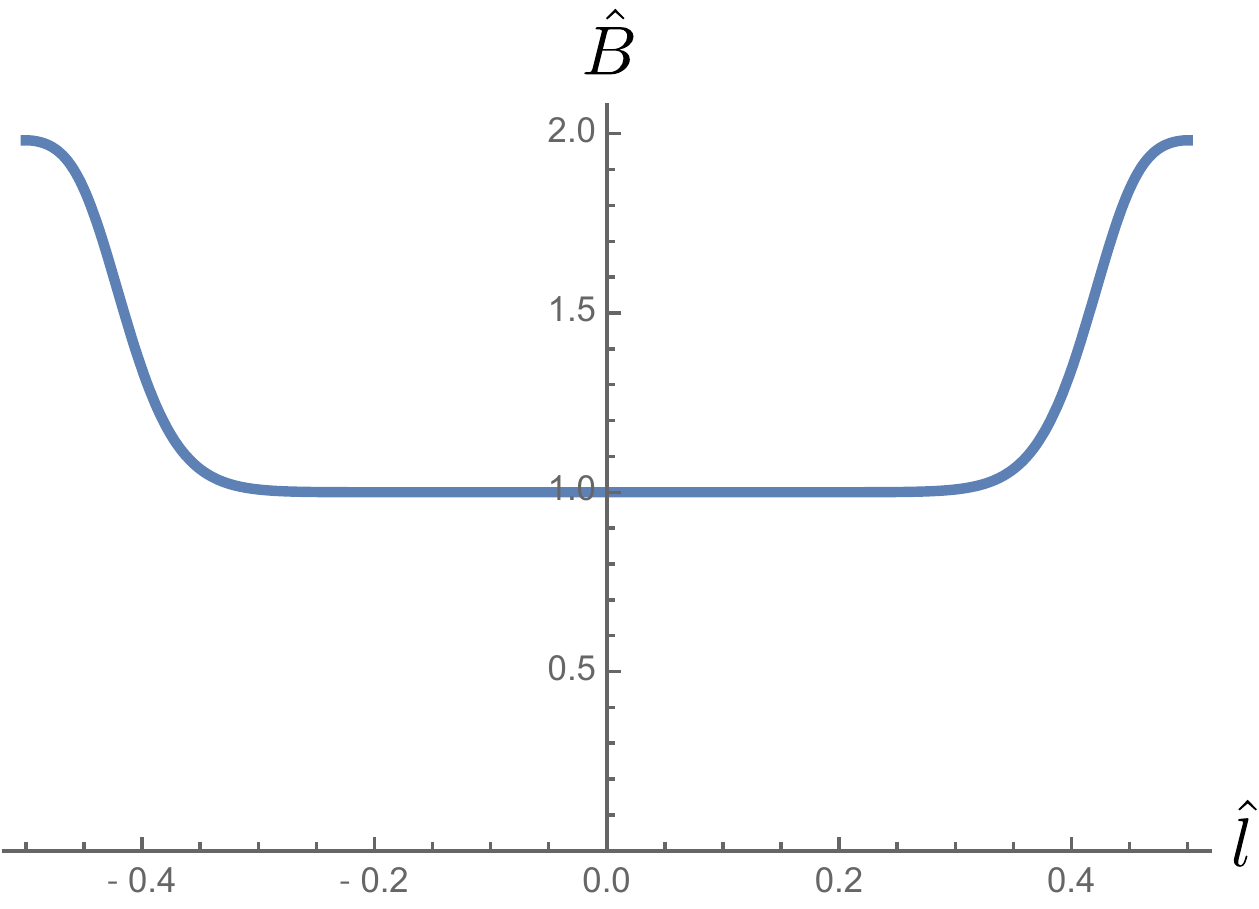}
        \caption{$\hat{B} = 2 - \exp[2\cos(\pi (2 \hat{l})^7)-2]$.}
        \label{well-sq-fig}
    \end{subfigure}%
    ~ 
    \begin{subfigure}[b]{0.5 \textwidth}
        \centering
	\includegraphics[width=1.0\textwidth]{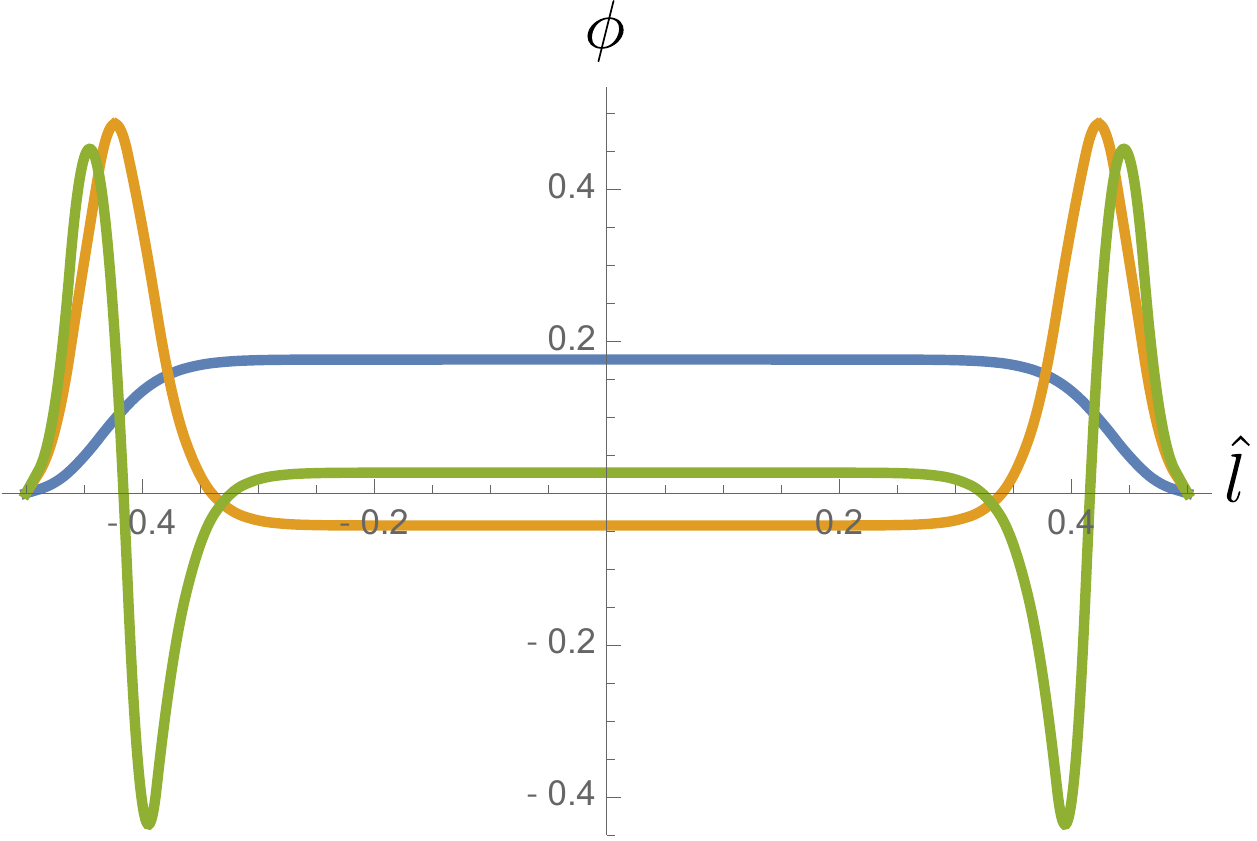}
        \caption{Eigenvalues: $1.37, 0.22, 0.15$}
	\label{eigenmode-sq-fig}
    \end{subfigure}

    \caption{Examples of numerical solutions of Eq.~(\ref{integral eq2a}).  Magnetic well plotted ($\hat{B}(\hat{l}) =B(l)/B_{\rm min}$ for $\hat{l} \in [-1/2, 1/2]$) next to first three eigenmodes (corresponding to the largest eigenvalues $\nu_i$).}
\label{eigenmode-figs}%
\end{figure*}

In Fig.~(\ref{eigenmode-fig}), the eigenmode solutions are plotted corresponding to the sinusoidal magnetic well, Fig.~\ref{well-sin-fig}, with $\hat{\kappa} = 1/|\hat{l}-1|$. The resulting eigenmode asymmetry is consistent with the general arguments made above, namely that its amplitude is enhanced at negative $\hat{l}$, where $\hat{\kappa}(\hat{l})$ is smaller.  

A double well structure for $B(l)$, Fig.~\ref{well-db-fig}, results in the mode structures shown in Fig.~(\ref{eigenmode-db-fig}).  Here $\hat{\kappa} = 1$.  We can see the demonstration of other qualitative features that seem characteristic, namely the appearance of only even eigenmodes (corresponding to the non-zero eigenvalues), and eigenfunction nodes at the endpoints, where the eigenfunctions also have zero derivative.

In Fig.~(\ref{eigenmode-sq-fig}), a smooth approximation to a square well, Fig.~\ref{well-sq-fig}, has been used, demonstrating how the single mode found in the perfect square well splits into a spectrum of modes that have structure near the well endpoints, where $B(l)$ is non-constant (here also $\hat{\kappa} = 1$).

\section{Conclusion}

In this paper, we have demonstrated the existence of an ion-driven trapped electron mode, which can exist in maximum-$J$ devices despite a stabilising electron magnetic drift.  Although not providing free energy to the mode, trapped electrons are nevertheless required, because they reduce the phase velocity sufficiently to allow for resonance with the ion magnetic drift.

The key results are given by Eqs.~(\ref{dispersion eqn1}), (\ref{integral eq2a}) and (\ref{dispersion eq2}).  Several important conclusions arise from inspecting these equations.  They demonstrate that the electron magnetic drift will be stabilising when the electrons are subjected to ``good'' bounce-averaged curvature, and even more strongly stabilising with an electron temperature gradient, $\eta_e > 0$.  The necessary balance between the ion and trapped-electron diamagnetic drifts is a feature shared with the ``ubiquitous mode" of \cite{Coppi-1977}; the balance must occur at some value of $k_\perp$, and thus this existence condition is guaranteed to be satisfied.  Furthermore, we note that the case where the balance is exact will be the zero-crossing point of the real part of the mode frequency in $k$-space, as is observed in numerical simulations \citep{Proll-2013}, where the transition occurs between ion and electron directed mode propagation.

We find that a simple integral equation, Eq.~(\ref{integral eq2a}), independent of mode frequency (and also independent of diamagnetic and magnetic drift frequencies) determines the mode structure (in the limit $\omega_{\ast} \gg \omega_{d}$).  This leads to the important conclusion that the mode structure will not necessarily peak in regions of bad curvature, and so some degree of favourable averaging should be expected to limit the overall instability of the ITEM.  This could explain the relatively small growth rates that have been previously observed in numerical simulations \citep{Proll-2013}.  

Solving Eq.~(\ref{integral eq2a}) in a simple analytically tractable limit illustrates that, for certain mode wavenumbers, the ITEM may be stabilized completely by favourable averaging of the ion magnetic drift, but it is not apparent how all wavenumbers might be stabilised for realistic magnetic configurations.  It is however noted that the mode may, for all wavenumbers, be particularly weak when the magnetic drift varies in an oscillatory manner on a scale smaller than that of the variation of $k_\perp$.  Numerical solutions of Eq.~(\ref{integral eq2a}) confirm qualitative properties of the solution, and demonstrate the possibility of further linear optimisation studies by numerical means.

\appendix
\section{Expressions for $f_1$ and $g_1$}\label{appA}

Here we provide explicit expressions for the functions $f_1$ and $g_1$:

\bn
f_1(\omega,l) = 1 + \frac{T_e}{T_i} \left[ 1 - h_1(\omega, l)\right]
\en
where

\begin{align}
h_1(\omega,l) = &\Gamma_0 \biggl[ 1 - \frac{\omega_{*i1}}{\omega} - \frac{(1 + \eta_i) \omega_{*i} \hat \omega_{di}}{\omega^2}\nonumber\\
	&+ b_0 \left( \frac{\eta_i \omega_{*i1}}{\omega} + \left( 2 \eta_i + \frac{1}{2} \right) \frac{\omega_{*i} \hat \omega_{di}}{\omega^2} \right) -b_0^2 \frac{\eta_i \omega_{*i} \hat \omega_{di}}{\omega^2} \biggr] \nonumber\\
	&-b_0 \Gamma_1 \left[\frac{\eta_i \omega_{*i1}}{\omega} + \frac{\omega_{*i} \hat \omega_{di}}{2 \omega^2}
	+ \left( \frac{3}{2} - b_0 \right) \frac{\eta_i \omega_{*i} \hat \omega_{di}}{\omega^2} \right]\nonumber\\	
	&+ b_1 \frac{\omega_{*i}}{\omega} \left[(1 + \eta_i)(\Gamma_0 - \Gamma_1) - \frac{b_0 \eta_i}{2}\left(3 \Gamma_0 - 4\Gamma_1 + \Gamma_2\right)  \right ],\label{appendix-h1-eqn}
\end{align}
and

\bn
g_1(\omega,\lambda) =\frac{1}{2} \left[1 - \frac{\omega_{*e1}}{\omega} 
	- \frac{3 (1+\eta_e) \omega_{*e} \tilde \omega_{de}}{2\omega^2}  \right],
\en
where $\omega_{*a1} = (T_a k_{\alpha 1}/e_a) d \ln n_a / d \psi$, $b_1 = 2 ({\boldsymbol k}_{\perp 0}\cdot {\boldsymbol k}_{\perp 1}) m_i T_i / (eB)^2$, and all other functions of ${\boldsymbol k}_\perp$ are understood to be evaluated at $k_{\alpha 0}$ and $k_{\psi 0}$.  Note that we have used $\Gamma_0' = \Gamma_1 - \Gamma_0$ and $\Gamma_1' = -\Gamma_1 + (\Gamma_0 + \Gamma_2)/2$ to perform the expansions, {\em i.e.} $\Gamma_0(b) \approx \Gamma_0(b_0) + b_1 \Gamma_0'$, {\em etc.}, yielding the terms of the final line of Eq.~(\ref{appendix-h1-eqn}).

\end{document}